\documentclass[preprint,aps,prl,floats,amsfonts,amssymb,graphics,draft]{revtex4}

\newcommand{\D}{{\rm d}}
\newcommand{\I}{{\rm i}}

\newcommand{\pdiff}[2]{{\frac{\partial #1}{\partial #2}}}

\newcommand{\bra}[1]{{\left\langle{#1}\right|}}
\newcommand{\ket}[1]{{\left|{#1}\right\rangle}}

\newcommand{\bracket}[2]{{\langle#1|#2\rangle}} 
\begin{document}
\title{Quantum Computation with Vibrationally Excited Molecules}
\author{Carmen M. Tesch\thanks{Carmen.Tesch@mpq.mpg.de} 
  and Regina de Vivie-Riedle\thanks{rdv@mpq.mpg.de}}
\address{MPI f{\"u}r Quantenoptik, 85741 Garching, Germany}
\date{\today}


\begin{abstract}
A new physical implementation for quantum computation 
is proposed. The  vibrational modes of molecules are
used to encode qubit systems. Global quantum logic gates
are realized using shaped femtosecond laser pulses
which are calculated
applying optimal control theory. 
The scaling of the system is favorable, sources for decoherence
can be eliminated. A complete set of one and two quantum gates is 
presented for a specific molecule.
Detailed analysis regarding experimental realization
shows that the structural resolution of today's pulse shapers
is easily sufficient for pulse formation. 
\end{abstract}

\maketitle

The possibilites offered through quantum computation have been well known 
for almost twenty years \cite{Feynman82,Deutsch85}.
In the last few years the first basic ideas for realization have 
been discussed and experimental efforts have been made using
cavity quantum electrodynamics \cite{Brune94}, trapped ions
\cite{Cirac95,Lange2000}, and nuclear magnetic resonance
\cite{Jones98,Glaser2000}.

In this Letter we demonstrate how to use 
vibrational modes of molecules in an ensemble
for complete quantum computation processes.
In an N-atomic molecule 3N-5 (or 3N-6) normal modes
can be identified, and each of these normal modes can be
used to define a quantum bit. Two different excitations in each
mode are referred to as $\ket{0}$ and $\ket{1}$.  
To implement quantum logic operations in this qubit system,
shaped femtosecond laser pulses in the IR-regime are used.
They are designed with
the help of optimal control theory (OCT) \cite{Tannor85,Judson92} which are in principle
realizable in pulse shaping experiments \cite{Witte,Kaindl}. Interaction
between different qubits is an inherent property
of the whole qubit system belonging to the same molecule.
For designing globally applicable laser pulses in just one qubit,
this interaction has to be suppressed; in the case of conditional 
gates this influence is explicitely used. Both cases can be realized
in our OCT-functional.

For initial state preparation again shaped
femtosecond laser pulses are used. They can be calculated with our
standard OCT-algorithm which can also compensate
for rotations of molecules in the gas phase.
\cite{Tesch2000,SundermannJCP}. Vibrational state preparation is optionally possible 
for the vibrationless ground state, fundamentals, overtones and combination modes.
In our work the initial state preparation is a preparation of a combination mode
for which maximally  two shaped femtosecond laser pulses of different
wavelengths are needed.

The detection of states after a computation can be achieved by applying
standard laser diagnostics (for example time resolved IR-spectroscopy \cite{Diller99},
laser induced fluorescence spectroscopy [LIF] \cite{Crim},
stimulated emission pumping [SEP] or resonant multi photon ionization [REMPI] \cite{Rosenwaks98}).

In the following part of the Letter 
we will show that vibrationally excited molecules
provide a realistic physical system
to implement a quantum computer.
We will demonstrate 
a realization of the above 
ideas using acetylene ($C_2H_2$) as a two-qubit model system. 
Our new OCT-functional, laser pulses for elementary global one and two quantum gates 
(which form a universal quantum gate altogether \cite{Sleator95}), 
and the preparation of entanglement will be presented.
An analysis of the shaped femtosecond laser pulses
will show that these pulses are simply structured.
The problem of decoherence is discussed 
and a solution is found. Error detection
is also feasible.


OCT is a very powerful tool for calculating laser pulses
guiding a quantum system to any selected objective,
in particular the preparation of eigenstates is enabled.
In this case the goal in OCT is to find a laser field
$\varepsilon(t)$, which drives a system from initial states
$\psi_{ik}(0)=\phi_{ik}$ at time $t=0$ to final target states $\phi_{fk}$ at a
fixed time $t=T$. The initial and target states correspond with the eigenstates 
defining our qubits. The calculated laser pulses  $\varepsilon(t)$
represent the quantum gates. 

Algorithmic schemes allow the formulation of the optimization
problem in terms of the maximization of the functional
\begin{eqnarray*}
K\left(\psi_{ik}(t),\psi_{fk}(t),\varepsilon(t)\right)&=&
\sum_k \left|\bracket{\psi_{ik}(T)}{\phi_{fk}}\right| ^2 -\int_0^T
\alpha|\varepsilon(t)| ^2\D t\\
&&-2\Re\left[\bracket{\psi_{ik}(T)}{\phi_{fk}}\int_0^T\bra{\psi_{fk}(t)}\left[\frac{\I}
    {\hslash}\left[H_0-\mu\varepsilon(t)\right]+\pdiff{}{t} \right]\ket{\psi_{ik}(t)}\D t
\right]
\end{eqnarray*}

The first term gives the overlap between the laser driven wavefunction
$\psi_{ik}(t)$
and the desired target states $\phi_{fk}$ ; the second one represents the laser field  $\varepsilon(t)$
which drives the system wavefunctions towards the target states. The laser field intensity
is limited by the penalty factor $\alpha$. 
The last term ensures that the time dependent Schr{\"o}dinger equation is fulfilled.
$H_0$ and the dipole moment vector field $\mu$ have been calculated $ab~initio$,
which means that all molecular characteristics of the electronic and nuclear movement
and their interactions are exactly implemented in the calculations. 

Varying the above functional with respect to the variables $\psi_{ik}(t)$,
$\psi_{fk}(t)$, and $\varepsilon(t)$ and searching for $\delta K=0$ leads
to a set of 2$k$+1 coupled differential equations.
The solution is found iteratively and results in the
optimal quantum gate $\varepsilon(t)$ as a self consistent solution to this
system.

The new feature which is of highest importance in the context of quantum computation
is the fact that we are able to optimize
$k$ transitions simultaneously with the above new functional.
Its recent implementation in our algorithms
enables us to calculate globally applicable quantum gates,
because for each basis state of our qubit system
the correct transition $k$ is optimized, and undesired transitions
can be explicitely suppressed.
For example in a two-qubit system a global NOT-pulse in the second qubit
has to fulfill $\ket{00}$~\begin{math}\to\end{math}~$\ket{01}$, 
$\ket{01}$~\begin{math}\to\end{math}~$\ket{00}$,
$\ket{10}$~\begin{math}\to\end{math}~$\ket{11}$ 
and $\ket{11}$~\begin{math}\to\end{math}~$\ket{10}$ at the same time.


These basic principles for concrete quantum computation
with vibrational modes are now discussed 
for an example system: acetylene.
$C_2H_2$ contains five normal modes ($n_1 n_2 n_3 n_4 n_5$) 
in the electronic ground state
($n_4$ and $n_5$ are doubly degenerate).
The asymmetric CH-stretching ($n_3$) and the $cis$-bending mode ($n_5$)
are IR-active.
We have chosen them to define our two-qubit system \cite{TeschCPL}
because direct experimental access is possible.
One quantum in each mode is referred to as $\ket{0}$, two quanta as $\ket{1}$
(other classifications would also have been possible). 
Our complete two-qubit system is described with the following four basis states:
$\ket{00}$~:=~($00101$); $\ket{10}$~:=~($00201$); 
$\ket{01}$~:=~($00102$); $\ket{11}$~:=~($00202$).

We have computed a two dimensional $ab~initio$
potential energy surface (PES) using a standard quantum chemical
package (GAUSSIAN98 \cite{Gaussian}).
One system-adapted coordinate is sufficient for
the description of one IR-active mode, because the relative distance between the two hydrogen
atoms and the relative distance between the two carbon atoms rests constant during the motion
(coordinates: $R \to$ $cis$-bending mode,
$d \to$ asymmetric CH-stretching mode, see \cite{Tesch2000}). 
In order to calculate the vibrational eigenstates $\phi_{fk}$ we solved the nuclear 
Schr{\"o}dinger equation for our PES on a grid applying a relaxation 
method followed by an additional
diagonalization step \cite{Tal-Ezer86,Neuhauser90}.
We calculated 80 eigenfunctions up to an energy of 23150 cm$^{-1}$
(the minimum energy is set to zero), four of them define our
qubit basis.
For the laser-matter-interaction we have computed the
$ab~initio$ dipole moment vector field, which can be decomposed into two parts,
each part being responsible for the excitation of one IR-active mode.
This decomposition and the energetic spacing between the two vibrations
ensure a very specific excitation of each mode
($cis$-bending mode 727~cm$^{-1}$ (13.75~$\mu$m), asymmetric CH-stretching mode 3289~cm$^{-1}$ (3.04~$\mu$m)).
For the construction of our quantum gate  $\varepsilon(t)$
the set of coupled differential equations is solved quantum dynamically applying
the split operator scheme \cite{Leforestier90} for the wave packet propagations.
  
With our new functional we  have constructed global laser pulses for the following
elementary one-quantum gates:
identity operation~\openone ~(trivial case or explicitely computed), bit flips (NOT), phase shifts of~$\pi$
and the Hadamard transformation in the second qubit 
(the case for the first qubit is analogous \cite{Tesch2002}):
\begin{eqnarray*}
\mbox{\openone:}\qquad\ket{0}\bra{0}+ \ket{1}\bra{1}\\
\mbox{NOT:}\qquad\ket{0}\bra{1}+ \ket{1}\bra{0}\\
\mbox{$\Pi$:}\qquad\ket{0}\bra{0}- \ket{1}\bra{1}
\end{eqnarray*}
\begin{eqnarray*}
\mbox{Hadamard}\qquad H\ket{0}\rightarrow \frac{1}{\sqrt2}\left[\ket{0}+\ket{1}\right]\quad\mbox{and}\quad
H\ket{1}\rightarrow \frac{1}{\sqrt2}\left[\ket{0}-\ket{1}\right]
\end{eqnarray*}
As a conditional two-quantum gate we have calculated a controlled NOT (CNOT)
with the first qubit acting as the control state:
\begin{eqnarray*}
\ket{00} \to \ket{00}\\
\ket{01} \to \ket{01}\\
\ket{10} \to \ket{11}\\
\ket{11} \to \ket{10}
\end{eqnarray*}

Exemplarily the optimized laser field for the CNOT gate is presented in Fig.~\ref{fig:CNOT}.
The first part of the figure displays the quantum gate $\varepsilon(t)$ versus time,
the second one depicts a fast Fourier transform to analyze the spectral width 
and components of the laser pulse.
It is not Fourier limited and pulse shaping is needed to form 
this electric field with a broad spectral range.
The last part of Fig.~\ref{fig:CNOT} analyzes the pulse structure in the time and frequency domain
simultaneously mirroring the temporal ordering of the frequency components. 
The substructure is significant and rather symmetric in time.

For the realization of all our quantum gates our method needs only one single shaped laser pulse;
a preceding special state preparation or pulse sequence is not necessary as in other approaches
like cavity QED, trapped ions and also NMR experiments. 
We have analyzed the mechanism according to which our quantum gates work. In all cases not
only states belonging to  our qubit system, but temporarily energetically higher and lower lying
vibrational levels are populated following a ladder climbing scheme. An intermediate coherent state
is built up before all population is transferred into the desired eigenstates. 
These eigenstates are stable, no refocusing pulses are needed as in NMR quantum computing. 
 
The ratios $r$ (defined as
\begin{math} \left|\bracket{\psi_{ik}(T)}{\phi_{fk}}\right| ^2\end{math})
we can reach are:
$r$(NOT)~$\ge$~91\%; $r$($\pi$)~$\ge$~97\%; $r$(Hadamard)~$\ge$~90\%; $r$(CNOT)~$\ge$~90\%.
(For comparison, in cases of state preparation with $k$~=~1
we can always reach efficiencies of $\ge$~98\%.)
In each case the laser pulse found by the OCT-algorithm
consists of several structural units, and the total laser matter
interaction time is about 700~fs.
The range of the frequency width is between 200~cm$^{-1}$ ($\pi$-pulse) and 
560~cm$^{-1}$ (CNOT-pulse).
The strength of the electric field $e$
is moderate except for the NOT-operation, but still below the ionization limit 
(maximum strength: $e$(NOT)~=~0.0224~a.u. (7.146$\cdot10^{13}$~W/cm$^{2}$);
$e$($\pi$)~=~0.0057~a.u. (1.1597$\cdot10^{12}$~W/cm$^{2}$); 
$e$(Hadamard)~=~0.0095~a.u.(3.226$\cdot10^{12}$~W/cm$^{2}$);
$e$(CNOT)~=~0.0070~a.u.(1.749$\cdot10^{12}$~W/cm$^{2}$)).
First investigations in other systems show that the field strengths 
can be lowered down \cite{Ulli,Ulli-dipl}.

With our OCT-algorithm we are also able to calculate
pulses which create entangled states.
The efficiency or fidelity of our Bell state preparation
(starting at one of our basis states and proceeding
via a Hadamard-state) is $r$~$\ge$~96\% in each case.
One part of the preparation of 
\begin{math}\frac{1}{\sqrt2}\left[\ket{00}+\ket{11}\right]\end{math}
is shown in Fig.~\ref{fig:BELL}.

The prepared Bell state is not a long-lived normal mode,
but a superposition state. In that context it is relevant to
investigate the time evolution of the corresponding wave function.
In Fig.~\ref{fig:BELL} and \ref{fig:HADAMARD}
we have depicted the time evolution
of the real part of our wave packets after the preparation
of a Bell state and a Hadamard state (of course, we have 
experimental access only to \begin{math} \left|\psi\right| ^2\end{math}).
The wave functions (real and imaginary part) develop 
in time systematically,
and \begin{math} \left|\psi\right| ^2\end{math} comes back in
predictable intervals. Their evolution in space is confined to
a certain area. Once again one does not need refocusing pulses.

To realize our calculated gates,
pulse shaping techniques are needed.
In the spectral range from 430~nm to
1.6~$\mu$m liquid crystal modulators (LCM) are
already commercially available, having typically 128 discrete
pixels. Powerful lasers operating in the mid-infrared 
already exist (free-electron laser FELIX (4-250~$\mu$m) \cite{vanHelden99}).
Pulse shapers that work in the mid-IR and far-IR spectral ranges
are actually under development; the development
of new types of nonlinear crystals already shows
promising results in those directions \cite{Rabitz2000}.
Anyhow, a realization of a pulse shaper
in the mid-IR and far-IR spectral ranges  
will follow a pixel scheme.
The mask function applied to the pulse shaper in order to generate the shaped
laser pulses is the direct interface
between theory and experiment.
Therefore we calculated the mask
functions needed to realize our quantum gates and investigated their complexity. 
The structure of the mask function (transmission and phase) is
very simple and only a few pixels are needed.
As an example the mask function for the CNOT is displayed in Fig.~\ref{fig:MASK}.   
The original pulse to be sent through the pulse shaper (Fig.~\ref{fig:MASK} top) is 
a 171~fs strictly Fourier limited  pulse (center frequency at
674~cm$^{-1}$, FWHM~590~cm$^{-1}$, maximum electric field strength 0.0280~a.u. (2.799$\cdot10^{13}$~W/cm$^{2}$)).

The ratios we can achieve with our gates at this stage are rather high,
but still below 100\%. 
We have analyzed the losses. Most result from leakage
into vibrations that lie outside our chosen Hilbert space and
therefore bear little disturbing influence for further calculations.
The effect is a diminishing signal when detecting results after a quantum computation process.
An estimation shows that at room temperature, normal pressure and an assumed
laser focus zone of 200~$\mu m$ in diameter and 1~$cm$ in length a few hundred
quantum gates could be applied successively. This is already enough for some very useful devices 
(e.g. a quantum factoring engine). 
We have found that the losses are localized in very few 
specific vibrational states
and can be detected there without disturbing the information
stored in our qubit system. Thus error detection is possible
in our proposed quantum computing model. 

The context of leakage is closely connected to the topic of decoherence.
One could think of two different sources of decoherence: 
resonances between different vibrations (anharmonic resonances and 
Coriolis coupling) and collisions between molecules.
Recent investigations have shown that anharmonic resonances do not cause 
decoherence at all, but sometimes provide easier mechanisms for designing quantum gates 
\cite{Ulli,Ulli-dipl}.
Regarding molecules in the gas phase, 
the number of collisions can be kept low, and typical lifetimes
of vibrations
are in the nanosecond regime or longer, depending on the experimental
conditions. The more significant
problem is motion of the molecules within the laser-matter-interaction
volume and possibly out of that zone before the next laser pulse
is applied. The repetition rate of the lasers ist too small and diminishes the
number of possible steps in a quantum computation process significantly. 
These problems can be avoided if fixed molecules are used. One can also think of
specially designed macro-molecules adsorbed on a surface with
the active centers still free to vibrate \cite{Rieder96,Mintova2001,Metzger2001}
or of molecules embedded in films \cite{Diller02}.
Molecular design
techniques for self assembling molecules are well developed 
at the frontier of nano technology \cite{Mena-Osteritz2000,Gimzewski99}.   
Rotations are also suppressed in fixed molecules, and therefore 
decoherence due to Coriolis coupling is precluded.

In the future, the design of macro-molecules consisting
of repeated subunits, e.g. subsystems connected through
conjugated CC-bonds, might play an important role for
molecular quantum computing processes. The interaction within each
subunit and between the subunits is secured using the
system-inherent influence of the vibrational modes upon one another
as explained above. In particular, a collective vibrational mode
of all subunits could be used as a data bus in such a quantum network. 
Every subunit and each vibrational mode has access 
to the same information coded in the excitation
of this collective vibration.
An N-atomic macro-molecule is an immense 
qubit system: the number of utilizable qubits
scales with 3N, in NMR and ion trap experiments with N.

In summary, vibrationally excited molecules are well suited
to implement quantum computation processes. 
Shaped femtosecond laser pulses act as extremely fast quantum gates and
can be designed by applying optimal conrol theory.
The calculated laser pulses are simply structured
and could be realized in an experimental surrounding where
decoherence is negligible.

\clearpage


\section*{Figure captions}

\begin{figure}[h]

  \caption[Controlled NOT gate]{Controlled NOT gate for a two-qubit system in vibrationally
excited acetylene. Depicted are the strength of the electric field, a fast Fourier transformation,
and a FROG-representation. The efficiency of the gate is $\ge$ 90\%.}
  \label{fig:CNOT}
\end{figure}

\begin{figure}[h]

  \caption[Preparation of a Bell State]{Preparation of a Bell state: $\frac{1}{\sqrt2}\left[\ket{00}+\ket{10}\right]
 \rightarrow \frac{1}{\sqrt2}\left[\ket{00}+\ket{11}\right]$, efficiency $\ge$ 98\%.
$R$ is the coordinate of the $cis$-bending mode, $d$ describes the asymmetric CH-stretching mode. 
Displayed are the laser pulse and the time evolution of the real part of the wave function during the pulse and afterwards.
The * indicates the initial and target state, the box labels states with the same \begin{math} \left|\psi\right| ^2\end{math}.}
  \label{fig:BELL}
\end{figure}

\begin{figure}[h]

  \caption[Preparation of a Hadamard State]{Preparation of a Hadamard state: $\ket{00}
 \rightarrow \frac{1}{\sqrt2}\left[\ket{00}+\ket{01}\right]$, efficiency $\ge$ 99\%.
$R$ is the coordinate of the $cis$-bending mode, $d$ describes the asymmetric CH-stretching mode.  
Displayed are the laser pulse and the time evolution of the real part of the wave function during the pulse and afterwards.
The * indicates the initial and target state, the box labels states with the same \begin{math} \left|\psi\right| ^2\end{math}.}
  \label{fig:HADAMARD}
\end{figure}

\begin{figure}[h]

  \caption[Mask function for the CNOT-gate]{Mask function for the CNOT-gate. For explanation see text.}
  \label{fig:MASK}
\end{figure}


\begin{thebibliography}{References}

\bibitem{Feynman82}
  R.~Feynman,
  Int.~J.~Theor.~Phys.~\textbf{21}, 467 (1982).

\bibitem{Deutsch85}
  D.~Deutsch,
  Proc.~R.~Soc.~London A \textbf{400}, 97 (1985).

\bibitem{Brune94}
  M.~Brune, F.~Schmidt-Kaler, A.~Maali, J.~Dreyer, E.~Hagley,
  J.M.~Raimond, and S.~Haroche,
  Phys. Rev. Lett \textbf{76}, 1800 (1996).

\bibitem{Cirac95}
  J.~I.~Cirac and P.~Zoller,
  Phys. Rev. Lett. \textbf{74}, 4091 (1995).

\bibitem{Lange2000}
  H.~C.~N{\"a}gerl, F.~Schmidt-Kaler, J.~Eschner, R.~Blatt,
  W.~Lange, H.~Baldauf, H.~Walther,
  \textsl{Linear ion traps for quantum information} in
  \textsl{The physics of quantum information},
  edited by D.~Bouwmeester, A.~Ekert, A.~Zeilinger,
  Springer-Verlag~Berlin~Heidelberg~2000, pp.~163-176.

\bibitem{Jones98}
  J.~A.~Jones and M.~Mosca,
  J. Chem. Phys. \textbf{109}, 1648 (1998).

\bibitem{Glaser2000}
  R.~Marx, A.~F.~Fahmy, J.~M.~Myers, W.~Bermel, S.~J.~Glaser,
  Phys. Rev. A \textbf{62}, 2000 012310-1-8. 

\bibitem{Tannor85}
  D. J. Tannor, S.~A. Rice,
  J. Chem. Phys. \textbf{83}, 5013 (1985).

\bibitem{Judson92}
  R. S. Judson, H.~Rabitz,
  Phys. Rev. Lett. \textbf{68}, 1500 (1992).

\bibitem{Witte}
  T.~Witte, D.~Zeidler, K.-L.~Kompa, D.~Proch, M.~Motzkus,
  Optics Lett. (2001), accepted.

\bibitem{Kaindl}
  R.~A.~Kaindl, M.~Wurm, K.~Reimann, P.~Hamm, A.~M.~Weiner, M.~Woerner,
  J.~Opt.~Soc.~Am.~B \textbf{17}, 2086 (2000).

\bibitem{Tesch2000}
  C.~M.~Tesch, K.-L.~Kompa, and R.~de~Vivie-Riedle,
  Chem. Phys. \textbf{267}, 173 (2001).  

\bibitem{SundermannJCP}
  K. Sundermann, R.~de~Vivie-Riedle,
  J. Chem. Phys. \textbf{110}, 1896 (1999).

\bibitem{Diller99}
  R.~Dziewior, K.Romey, R.~Diller,
  Laser Chem. \textbf{19}, 173 (1999).

\bibitem{Crim}
  A.~L.~Utz, E.~Carrasquillo~M., J.~D.~Tobiason, F.~F.~Crim,
  Chem. Phys. \textbf{190}, 311 (1995).

\bibitem{Rosenwaks98}
  T. Arusi-Parpar, R.~P.~Schmid, Y.~Ganot, I.~Bar, S.~Rosenwaks,
  Chem.~Phys.~Lett. \textbf{287}, 347 (1998).

\bibitem{Sleator95}
  T.~Sleator and H.~Weinfurter,
  Phys. Rev. Lett. \textbf{74}, 4087 (1995).

\bibitem{TeschCPL}
  C.~M.~Tesch, L.~Kurtz, R.~de~Vivie-Riedle,
  Chem. Phys. Lett. \textbf{343}, 633 (2001).

\bibitem{Gaussian}
  Gaussian 98, Revision A.7,
  M.~J. Frisch et al.,
  Gaussian Inc., Pittsburgh PA, 1998.

\bibitem{Tal-Ezer86}
  H.~Tal-Ezer, R.~Kosloff,
  Chem. Phys. Lett. \textbf{127}, 223 (1986).

\bibitem{Neuhauser90}
  D. Neuhauser,
  J. Chem. Phys. \textbf{93}, 2611 (1990).

\bibitem{Leforestier90}
  C.~Leforestier, R.~H.~Bisseling, C.~Cerjan,
  M.~D.~Feit, R.~Friesner, A.~Guldberg, A.~Hammerich,
  G.~Jolicard, W.~Karrlein, H.-D.~Meyer, N.~Lipkin,
  O.~Roncero, R.~Kosloff,
  Journal of Computational Physics \textbf{94}, 59 (1991). 

\bibitem{Tesch2002}
  C.~M.~Tesch, R.~de~Vivie-Riedle,
  in preparation.

\bibitem{Ulli}
  U.~Troppmann, C.~M.~Tesch, R.~de~Vivie-Riedle,
  in preparation.

\bibitem{Ulli-dipl}
  U.~Troppmann, Diploma Thesis (2002).

\bibitem{vanHelden99}
  G.~van~Helden, I.~Holleman, M.~Putter,
  A.·~J.~A.~van~Roij, G.~Meijer,
  Chem.~Phys.~Lett. \textbf{299}, 171 (1999).

\bibitem{Rabitz2000}
  H.~Rabitz, R.~de~Vivie-Riedle, M.~Motzkus, K.-L.~Kompa,
  Science \textbf{288}, 824 (2000).

\bibitem{Rieder96}
  G.~Meyer, S.~Z\"{o}phel, K.-H-~Rieder,
  Appl. Phys. Lett \textbf{69} (21), 3185 (1996).

\bibitem{Mintova2001}
  S.~Mintova, T.~Bein,
  Advanced Materials \textbf{13} (24), 1880 (2001).

\bibitem{Metzger2001}
  T.~H.~Metzger, S.~Mintova, T.~Bein,
  Microporous and Mesoporous Materials \textbf{43}, 191 (2001).

\bibitem{Diller02}
  R.~Diller, J.~Herbst, K.~Heyne,
  AMOP Fr{\"u}hjahrstagung Osnabr{\"u}ck 2002, SYFS VIII.

\bibitem{Mena-Osteritz2000}
  E.~Mena-Osteritz, A.~Meyer, B.~M.~W.~Langeveld-Voss,
  R.~A.~J.~Janssen, E.~W.~Meyer, P.~B\"{a}uerle,
  Angew.~Chem.~\textbf{112}, 2792 (2000).

\bibitem{Gimzewski99}
  J.~K.~Gimzewski, C.~Joachim,
  Science \textbf{283}, 168 (1999).
\end{thebibliography}
\end{document}